\documentclass[prl,twocolumn,showpacs]{revtex4}
\usepackage{graphicx}

\begin{document}
\newcommand{\sgn}{\ensuremath\mathrm{sgn}}
\newcommand{\lone}{\ensuremath{\hat{\lambda}_1}}
\newcommand{\ltwo}{\ensuremath{\hat{\lambda}_2}}

\title{The Communication Cost of Simulating Bell Correlations}
\author{B. F. Toner} 
\email{toner@theory.caltech.edu} 
\author{D. Bacon} 
\email{dabacon@cs.caltech.edu} 
\affiliation{Institute for Quantum Information,
California Institute of Technology, Pasadena, CA 91125}
\affiliation{Department of Physics,
California Institute of Technology, Pasadena, CA 91125}

\begin{abstract} 
  What classical resources are required to simulate quantum
  correlations?  For the simplest and most important case of local
  projective measurements on an entangled Bell pair state, we show
  that exact simulation is possible using local hidden variables
  augmented by just one bit of classical communication.  Certain
  quantum teleportation experiments, which teleport a single qubit,
  therefore admit a local hidden variables model.
\end{abstract}

\pacs{03.65.Ud,03.65.Ta,03.67.-a,03.67.Hk,03.67.Dd}

\maketitle

Recent theoretical research into quantum algorithms~\cite{Shor:94a},
quantum communication complexity~\cite{Raz:99a}, and quantum
cryptography~\cite{Bennett:84a} has shown that quantum devices are
more powerful than their classical counterparts.  Indeed, the
flourishing field of quantum information theory~\cite{Nielsen:00a}
aims to provide an information-theoretic quantification of the power
underlying quantum resources.  One important feature of quantum theory
lies in the statistical correlations produced by measurements on local
components of a quantum system.  Almost forty years ago, John Bell
showed that such correlations cannot be explained by descriptions
based on realistic properties of local subsystems~\cite{Bell:64a}.  To
experimentally distinguish quantum correlations from those produced by
local hidden variables theories, Bell introduced the notion of {\em
  Bell inequalities}, with subsequent experimental evidence falling
squarely on the side of quantum theory~\cite{Aspect:82a}. Bell
inequalities, whilst usually considered relevant only to foundational
studies of quantum theory, answer a fundamental information-theoretic
question: what correlations can be produced between separate classical
subsystems, which have interacted in the past, if no communication
between the subsystems is allowed?  Violation of a Bell inequality,
however, does nothing to {\em quantify} what classical
information-processing resources are required to simulate a particular
set of quantum correlations.

The simplest and most important example of quantum correlations
involves the correlations produced by projective measurements on a
Bell pair.  Bell pairs are the maximally-entangled states of two
quantum bits (qubits) and are the basic resource currency of bipartite
quantum information theory.  Various equivalences are known: one
shared Bell pair plus two bits of classical communication can be used
to teleport one qubit~\cite{Bennett:93a} and, conversely, one shared
Bell pair plus a single qubit of communication can be used to send two
bits of classical communication via superdense
coding~\cite{Bennett:92a}.

Consider the gedanken experiment of Einstein, Podolsky, and
Rosen~\cite{Einstein:35a} (EPR), as reformulated by
Bohm~\cite{Bohm:51a}. Two spatially separate parties, Alice and Bob,
each have a spin-$\frac12$ particle, or qubit.  The global spin wave
function is the entangled Bell singlet state (also known as an EPR
pair) $|\psi\rangle={1 \over \sqrt{2}} \left(|\uparrow \rangle_A
  |\downarrow \rangle_B - |\downarrow \rangle_A |\uparrow_B \rangle
\right)$. The spin states $|\uparrow\rangle$, $|\downarrow\rangle$ are
defined with respect to a local set of coordinate axes:
$|\uparrow\rangle$ (resp. $|\downarrow \rangle$) corresponds to spin
along the local $+\hat{z}$ (resp.  $-\hat{z}$) direction.  Alice and
Bob each measure their particle's spin along a direction parameterized
by a three-dimensional unit vector: Alice measures along $\hat{a}$,
Bob along $\hat{b}$.  Alice and Bob obtain results,
$\alpha\in\{+1,-1\}$ and $\beta \in\{+1,-1\}$, respectively, which
indicate whether the spin was pointing along ($+1$) or opposite ($-1$)
the direction each party chose to measure.  Locally, Alice and Bob's
outcomes appear random, with expectation values $\langle \alpha
\rangle = \langle \beta \rangle =0$, but their joint probabilities are
correlated such that that $\langle \alpha \beta \rangle = - \hat{a}
\cdot \hat{b}$.  We refer to these correlations as {\em Bell
  correlations}.

It is not possible to reproduce these correlations using a protocol
which draws on random variables shared between Alice and Bob, but does
not allow communication after they have selected
measurements~\cite{Bell:64a}.  So how much communication {\em is}
required to exactly simulate
them~\cite{Maudlin:92a,Brassard:99a,Steiner:00a,Janos:02b,Bacon:02a,Cerf:00a,Masar:01a}?
Naively, Alice can just tell Bob the direction of her measurement
$\hat{a}$ (or vice versa) but this requires an infinite amount of
communication.  The question of whether a simulation can be done with
a finite amount of communication was raised independently by
Maudlin~\cite{Maudlin:92a}, Brassard, Cleve, and
Tapp~\cite{Brassard:99a}, and Steiner~\cite{Steiner:00a}.  Their
approaches differ in how the communication cost of the simulation is
defined: Brassard {\it et al.}  take the cost to be the number of bits
sent in the worst case; Steiner, the average.  (Steiner's model is
weaker because the amount of communication in the worst case can be
unbounded, although such cases occur with probability zero.)  Brassard
{\em et al.} present a protocol which simulates Bell correlations
using exactly eight bits of communication (since improved to six
bits~\cite{Janos:02b}).  Surprisingly~\cite{Bacon:02a}, the only lower
bound for the amount of communication is given by Bell's theorem: at
least some communication is needed.  Here we present a simple protocol
that uses just one bit of communication.

We first note three simple properties of Bell correlations: (i) if
$\hat a= \hat b$, then we must have $\alpha = - \beta$: Alice and Bob
must output perfectly anticorrelated bits; (ii) either party can
reverse their measurement axis and flip their output bit; and (iii)
the joint probability is only dependent on $\hat a$ and $\hat b$ via
the combination $\hat a \cdot \hat b$.  In his original paper, Bell
gave a local hidden variables model that reproduces these three
properties for all possible axes, but his model fails to reproduce
Bell correlations because the statistical correlations when $\hat a
\not = \hat b$ are not as strong as those of quantum
mechanics~\cite{Bell:64a}.  The protocol we describe below is inspired
by Bell's original protocol.  Property (iii) implies that we may
restrict attention to rotationally invariant protocols, for which all
probabilities depend only on $\hat{a} \cdot \hat{b}$ and not $\hat{a}$
and $\hat{b}$ separately, by {\em randomizing over all inputs with the
  same dot product}.  More precisely, suppose $P$ is any protocol that
simulates the correlations.  Then define a new protocol $P'$ whose
hidden variables consist of (i) those required by $P$, and (ii) a
random rotation $R \in \mathrm{SO}(3)$.  Protocol $P'$ then consists
of running protocol $P$ on $R \hat a$ and $R \hat b$ in place of $\hat
a$ and $\hat b$.

We now describe our protocol.  Alice and Bob share two random
variables $\lone$ and $\ltwo$ which are real three-dimensional unit
vectors. They are chosen independently and distributed uniformly over
the unit sphere.

\begin{figure}[htb]
\centering
\includegraphics{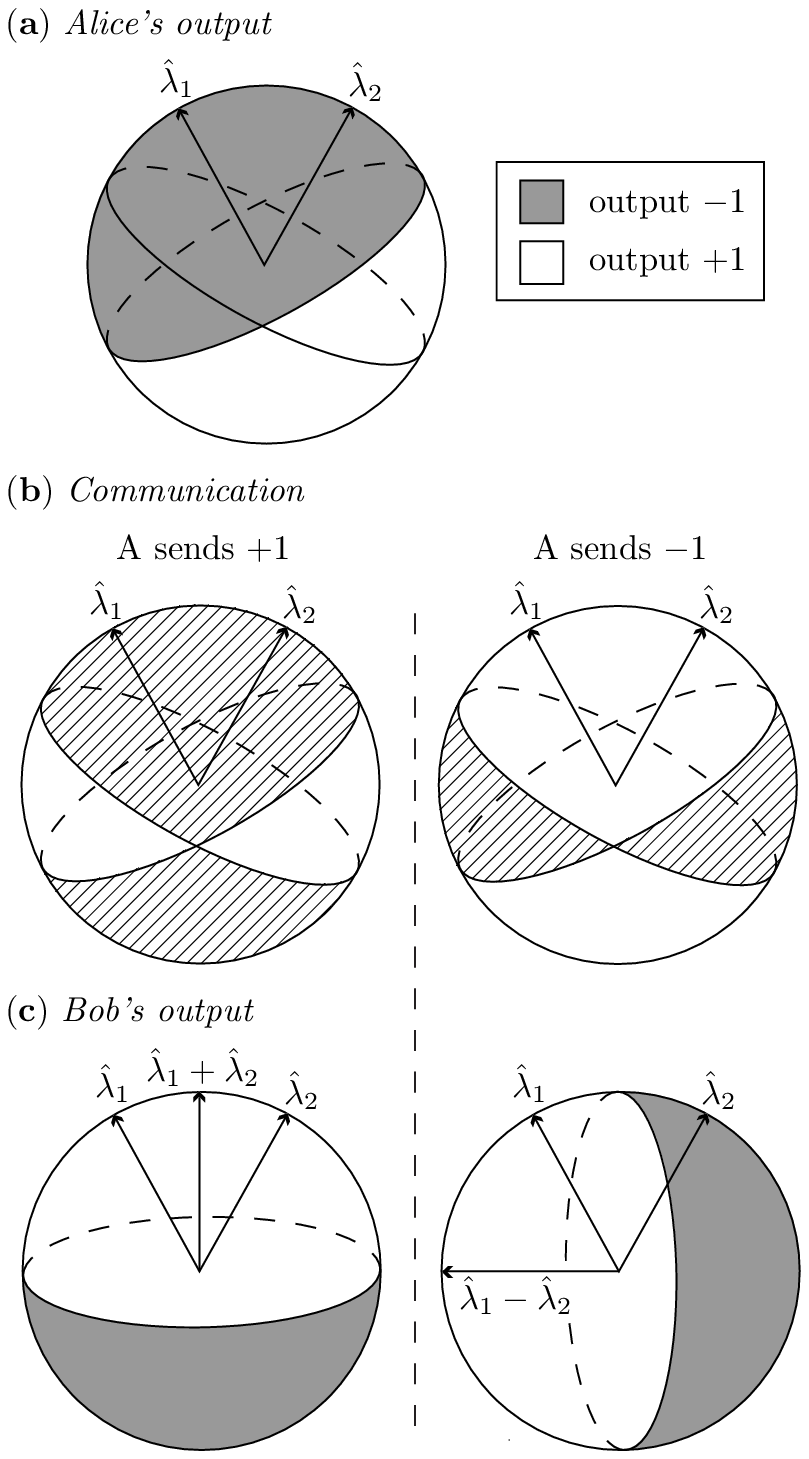}
\caption{{\em The protocol}: The shared unit vectors $\lone$ and
  $\ltwo$ described in the text divide the Bloch sphere into four
  quadrants, as shown.  Alice and Bob's actions depend on which
  quadrant their respective measurement axes lie in, and in Bob's
  case, the bit he receives from Alice.  ({\bf a}) {\em Alice's
    output}: if $\hat a$ lies in the shaded region, Alice outputs
  $-1$; in the unshaded region, she outputs $+1$. ({\bf b}) {\em The
    communication}: Alice sends $c = +1$ if her measurement axis lies
  in the N or S quadrants, and $-1$ otherwise.  ({\bf c}) {\em Bob's
    output}: this depends on the bit received from Alice.  The shading
  is as for ({\bf a}).  } \label{fig:1}
\end{figure}

The protocol proceeds as follows:
\begin{enumerate}
\item Alice outputs $\alpha=-\sgn (\hat{a} \cdot
  \lone)$.
  
\item Alice sends a single bit $c \in \{ -1, +1 \}$ to Bob where
  $c=\sgn (\hat{a} \cdot \lone) \sgn (\hat{a} \cdot
  \ltwo)$.
  
\item Bob outputs $\beta = \sgn \left[\hat b \cdot(\lone + c\ltwo)\right]$,
\end{enumerate}
where we have used the $\sgn$ function defined by ${\rm
  sgn}(x)=+1~{\rm if}~x \geq 0$ and $\sgn (x)=-1~{\rm if}~x<0$.  A
geometric description of our protocol is given in Fig.~\ref{fig:1}.
We note immediately that Bob obtains {\em no information} about
Alice's output from the communication.

\begin{figure*}[htb]
\centering
\includegraphics{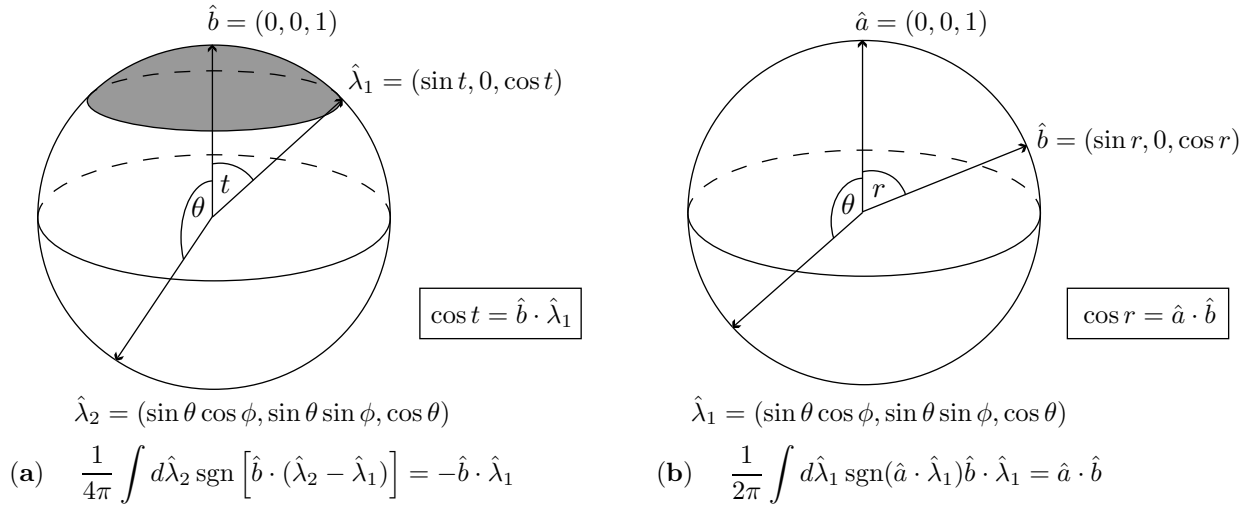}
\caption{{\em Construction used to evaluate 
    Eq.~(\ref{eq:ab})}: ({\bf a}) We first integrate over $\ltwo$,
  taking $\hat b$ to point along the positive
  $z$-axis~\cite{Schatten:93a}.  Observe that $\sgn \left[\hat b
    \cdot(\ltwo-\lone)\right]$ is positive in the top spherical cap
  (shaded) and negative otherwise. The area of the top spherical cap
  is $A_+ = 2\pi \int_0^t \sin \theta d \theta = 2\pi (1-\cos t)$
  where $\cos t = \hat b \cdot \lone$, hence $\int d \ltwo\, \sgn
  \left[\hat b \cdot(\ltwo-\lone)\right] = A_+ - (4\pi -A_+) = -4\pi
  \cos t = -4\pi \hat b \cdot \lone$.  ({\bf b}) We now take $\hat a$
  to point along the positive $z$-axis~\cite{Gisin:99a}, set $\hat b =
  (\sin r, 0, \cos r)$, and integrate over $\lone$, obtaining $\int d
  \lone\, \sgn(\hat a \cdot \lone) \hat b \cdot \lone = \int^\pi_{0}
  \sin \theta d \theta \int_0^{2 \pi} d \phi \;\sgn ( \cos \theta)
  \left( \cos \theta \cos r + \sin \theta \cos \phi \sin r \right) = 2
  \pi \cos r = 2 \pi \hat a \cdot \hat b.  $ } \label{fig:2}
\end{figure*}

We now prove the protocol reproduces the correct expectation values.
Each party's output changes sign under the symmetry $\lone
\leftrightarrow - \lone$, $\ltwo \leftrightarrow -\ltwo$, so $\langle
\alpha \rangle = \langle \beta \rangle =0$ because $\lone$ and $\ltwo$
are uniformly distributed.  The joint expectation value $\langle
\alpha \beta \rangle$ can be calculated using
\begin{eqnarray}
\label{eq:sum}
\langle \alpha \beta \rangle &=& E\Bigg\{-\sgn (\hat{a} \cdot
  \lone) \times  \nonumber \\ &&
\sum_{d=\pm 1} \frac{(1+cd)}{2}\,\, \sgn \left[\hat b
    \cdot(\lone + d \ltwo)\right] \Bigg\}
\end{eqnarray}
where $E\left\{ x \right\} ={1 \over (4\pi)^2} \int d \lone \int d
\ltwo \, x$, $c = \sgn (\hat{a} \cdot \lone)\, \sgn (\hat{a} \cdot
\ltwo)$ and we have used the trick that $(1 + cd)/2 = 1$ if $c=d$ and
$0$ if $c \not=d$.  After substituting for $c$ and expanding Eq.~(1),
we obtain the sum of four terms (because each term inside the
summation sign is itself the sum of two terms) and, using $\sgn
(\hat{a} \cdot \lone)\, c = \sgn (\hat{a} \cdot \ltwo)$, we note that
the four terms are related by the symmetries $\lone \leftrightarrow
\ltwo$ or $\ltwo \leftrightarrow -\ltwo$, so each has the same
expectation value.  Hence
  \begin{equation}
    \label{eq:ab}
\langle \alpha \beta \rangle = E\left\{2\, \sgn (\hat{a} \cdot
  \lone) \,\sgn \left[\hat b
    \cdot\left(\ltwo-\lone\right)\right] \right\}.    
  \end{equation}
This integral may be evaluated with the help of the two diagrams shown
in Fig.~\ref{fig:2}, with the result that $\langle \alpha \beta \rangle
= - \hat a \cdot \hat b$, as required.

Our protocol exactly simulates the quantum mechanical probability
distribution for projective measurements on the singlet Bell pair
state.  If a large number of simulations are performed in parallel,
the communication may be compressed.  To see this, assume Alice's
measurement vector $\hat a$ is uniformly distributed (if not, we
randomize, as outlined above).  Then, if $\lone \cdot \ltwo = \cos
\eta$, Alice sends $-1$ with probability $\eta/\pi$ and $1$ with
probability $1-\eta/\pi$, so that the communication can be compressed
to $\int_0^{\pi/2} \sin \eta\, d \eta H(\eta/\pi) \approx 0.85$ bits,
where $H(\eta/\pi)$ is the Shannon entropy.  This encoding depends on
the shared unit vectors $\lone$ and $\ltwo$: a third party without
access to the hidden variables will observe Alice sending uniformly
distributed bits to Bob.

Our protocol is easily modified to simulate joint measurements on any
maximally entangled state of two qubits, because every such state is
related to the singlet by a local change of basis and thus may be
simulated by rotating and/or reflecting the input vectors $\hat a$ and
$\hat b$, before running our protocol.  

Now consider the following experiment: Alice prepares a qubit in a
state unknown to Bob.  She then teleports the qubit to Bob, who
performs a projective measurement on it, along a direction unknown to
Alice.  We shall show that this experiment admits a local hidden
variables description.  We first note that quantum teleportation
experiments do not purport to test whether quantum mechanics allows a
local hidden variables model; rather they aim to distinguish quantum
teleportation from other protocols Alice and Bob might carry out using
classical communication, but no entanglement~\cite{Boschi:98a}.  From
this point of view, teleportation experiments represent
``investigations {\em within} quantum
mechanics''~\cite{Braunstein:01a}, rather than comparisons of quantum
mechanics with classical local hidden variables models~\cite{Hardy:99a}.  With this
distinction in mind, it is still interesting to ask whether
teleportation experiments {\em can} be explained by a local hidden
variables model.

If one allows an infinite amount of classical communication from Alice
to Bob, then there is a trivial local hidden variables model, for
Alice can just send a classical description of the state to Bob, who
then simulates his measurement.  We now give a local hidden variables
model that requires only two bits of communication, which is the {\em
  same} amount as the quantum teleportation protocol.  The
construction is based on Ref.~\cite{Cerf:00a}, where the procedure is
termed ``classical teleportation.''  It is sufficient to consider the
case where Alice prepares the qubit in a pure state, which we suppose
has spin aligned along the axis $\hat a$.  We suppose Bob's
measurement is aligned along the axis $\hat b$.  Alice and Bob share
uniformly distributed random three-dimensional unit vectors $\lone$
and $\ltwo$ (which can be thought of as hidden variables carried by
the Bell pair used for teleportation). The protocol is as follows:
\begin{enumerate}
\item Alice sends $c_1 = \sgn (\hat{a} \cdot
  \lone)$ and $c_2 = \sgn (\hat{a} \cdot
  \ltwo)$ to Bob.
\item Bob outputs $\beta = \sgn \left[\hat b \cdot\left( c_1 \lone + c_2 
  \ltwo \right)\right]$.
\end{enumerate}
It is easy to verify that $\langle \beta \rangle = \hat a \cdot \hat
b$, as required.  We also note that the two bits sent appear
completely random to a party without access to the hidden variables.

It is usual in teleportation experiments to have (i) a third party
Victor supply Alice with a quantum state unknown to her, and (ii) Bob
hand off the teleported state to Victor (or another party) to measure,
rather than measuring it himself.  Such a distinction is not important
for the question we address, because the qubit transmitted from Victor
to Alice, for example, can carry hidden variables describing its
state.  The point is that local hidden variables are {\em hidden}:
although it is convenient to describe a local hidden variables model
as if Alice and Bob had access to the hidden variables, the model
still exists even if the hidden variables are inaccessible to them.
There is no way for the experimenters to tell whether their experiment
is described by quantum theory or by ``gremlins'' within their
apparatus, executing the local hidden variables protocol described
above.

Are there quantum teleportation experiments which do not have such a
local hidden variables description?  One obvious possibility is an
experiment that teleports entanglement itself.  But there is a more
subtle possibility.  If we allow Bob to measure the qubit using
elements of a positive operator-valued measure, then there may not be
a local hidden variables description which respects the two bit
classical communication bound.  More generally, if Alice teleports $n$
qubits (which requires $2n$ bits of communication) and Bob makes a
joint measurements on them, then it is known that any exact local
hidden variables theory requires that Alice send at least a constant
times $2^n$ bits of communication in the worst
case~\cite{Brassard:99a}.  Whether this holds for protocols with
bounded error is an important open question.

Finally, using the classical teleportation protocol, we obtain a (not
necessarily optimal) protocol to simulate joint projective
measurements on partially entangled states of two qubits, which uses
two bits of communication: Alice first simulates her measurement and
determines the post-measurement state of Bob's qubit; Alice and Bob
then execute the classical teleportation protocol.

The results presented here offer an intriguing glimpse into the nature
of correlations produced in quantum theory.  If we interpret Bell
inequality violation to mean that some communication is necessary to
simulate Bell correlations, then our results prove that the minimal
amount, one bit, is all that is necessary for projective measurements
on Bell pairs.  Is our straightforward protocol an indication of a
deep structure in quantum correlations? We hope that our protocol and
the development of a general theory of the communication cost of
simulating quantum correlations will help shed light on this
fundamental question.

{\em Acknowledgements.}---We thank John Preskill, Andrew Doherty,
Patrick Hayden, Andre Methot, Carlos Mochon, and Michael Steiner for
useful suggestions.  This work was supported in part by the National
Science Foundation under grant EIA-0086038, through the Institute for
Quantum Information.


\begin{thebibliography}{24}
\expandafter\ifx\csname natexlab\endcsname\relax\def\natexlab#1{#1}\fi
\expandafter\ifx\csname bibnamefont\endcsname\relax
  \def\bibnamefont#1{#1}\fi
\expandafter\ifx\csname bibfnamefont\endcsname\relax
  \def\bibfnamefont#1{#1}\fi
\expandafter\ifx\csname citenamefont\endcsname\relax
  \def\citenamefont#1{#1}\fi
\expandafter\ifx\csname url\endcsname\relax
  \def\url#1{\texttt{#1}}\fi
\expandafter\ifx\csname urlprefix\endcsname\relax\def\urlprefix{URL }\fi
\providecommand{\bibinfo}[2]{#2}
\providecommand{\eprint}[2][]{\url{#2}}

\bibitem[{\citenamefont{Shor}(1994)}]{Shor:94a}
\bibinfo{author}{\bibfnamefont{P.~W.} \bibnamefont{Shor}}, in
  \emph{\bibinfo{booktitle}{Proceedings of the 35th Annual Symposium on the
  Foundations of Computer Science}}, edited by
  \bibinfo{editor}{\bibfnamefont{S.}~\bibnamefont{Goldwasser}}
  (\bibinfo{publisher}{IEEE Computer Society}, \bibinfo{address}{Los Alamitos,
  CA}, \bibinfo{year}{1994}), pp. \bibinfo{pages}{124--134}; \bibinfo{author}{\bibfnamefont{L.}~\bibnamefont{Grover}}, in
  \emph{\bibinfo{booktitle}{Proceedings of the 28th Annual ACM Symposium on the
  Theory of Computation}} (\bibinfo{publisher}{ACM Press}, New York,
  \bibinfo{year}{1996}), pp. \bibinfo{pages}{212--219}.

\bibitem[{\citenamefont{Raz}(1999)}]{Raz:99a}
\bibinfo{author}{\bibfnamefont{R.}~\bibnamefont{Raz}}, in
  \emph{\bibinfo{booktitle}{Proceedings of the 31st ACM Symposium on Theory of
  Computing}} (\bibinfo{publisher}{ACM Press}, New York, \bibinfo{year}{1999}), pp.
  \bibinfo{pages}{358--367}.

\bibitem[{\citenamefont{Bennett and Brassard}(1984)}]{Bennett:84a}
\bibinfo{author}{\bibfnamefont{C.~H.} \bibnamefont{Bennett}} \bibnamefont{and}
  \bibinfo{author}{\bibfnamefont{G.}~\bibnamefont{Brassard}}, in
  \emph{\bibinfo{booktitle}{Proceedings of IEEE International Conference on
  Computers, Systems, and Signal Processing}}
  (\bibinfo{publisher}{IEEE Computer Society}, 
  \bibinfo{address}{Los Alamitos, CA}, 
  \bibinfo{year}{1984}), pp. \bibinfo{pages}{175--179}.

\bibitem[{\citenamefont{Nielsen and Chuang}(2000)}]{Nielsen:00a}
\bibinfo{author}{\bibfnamefont{M.~A.} \bibnamefont{Nielsen}} \bibnamefont{and}
  \bibinfo{author}{\bibfnamefont{I.~L.} \bibnamefont{Chuang}},
  \emph{\bibinfo{title}{Quantum Computation and Quantum Information}}
  (\bibinfo{publisher}{Cambridge University Press}, \bibinfo{address}{New
  York}, \bibinfo{year}{2000}).

\bibitem[{\citenamefont{Bell}(1964)}]{Bell:64a}
\bibinfo{author}{\bibfnamefont{J.~S.} \bibnamefont{Bell}},
  \bibinfo{journal}{Physics} \textbf{\bibinfo{volume}{1}}, \bibinfo{pages}{195}
  (\bibinfo{year}{1964}).

\bibitem[{\citenamefont{Aspect et~al.}(1982{\natexlab{a}})\citenamefont{Aspect,
  Grangier, and Roger}}]{Aspect:82a}
\bibinfo{author}{\bibfnamefont{A.}~\bibnamefont{Aspect}},
  \bibinfo{author}{\bibfnamefont{P.}~\bibnamefont{Grangier}}, \bibnamefont{and}
  \bibinfo{author}{\bibfnamefont{G.}~\bibnamefont{Roger}},
  \bibinfo{journal}{Phys. Rev. Lett.} \textbf{\bibinfo{volume}{49}},
  \bibinfo{pages}{91} (\bibinfo{year}{1982}{\natexlab{a}}); \bibinfo{author}{\bibfnamefont{A.}~\bibnamefont{Aspect}},
  \bibinfo{author}{\bibfnamefont{J.}~\bibnamefont{Dalibard}}, \bibnamefont{and}
  \bibinfo{author}{\bibfnamefont{G.}~\bibnamefont{Roger}},
  \bibinfo{journal}{Phys. Rev. Lett.} \textbf{\bibinfo{volume}{49}},
  \bibinfo{pages}{1804} (\bibinfo{year}{1982}{\natexlab{b}}); \bibinfo{author}{\bibfnamefont{G.}~\bibnamefont{Weihs}},
  \bibinfo{author}{\bibfnamefont{T.}~\bibnamefont{Jennewein}},
  \bibinfo{author}{\bibfnamefont{C.}~\bibnamefont{Simon}},
  \bibinfo{author}{\bibfnamefont{H.}~\bibnamefont{Weinfurter}},
  \bibnamefont{and}
  \bibinfo{author}{\bibfnamefont{A.}~\bibnamefont{Zeilinger}},
  \bibinfo{journal}{Phys. Rev. Lett.} \textbf{\bibinfo{volume}{81}},
  \bibinfo{pages}{5039} (\bibinfo{year}{1998});  \bibinfo{author}{\bibfnamefont{W.}~\bibnamefont{Tittel}},
  \bibinfo{author}{\bibfnamefont{J.}~\bibnamefont{Brendel}},
  \bibinfo{author}{\bibfnamefont{H.}~\bibnamefont{Zbinden}}, \bibnamefont{and}
  \bibinfo{author}{\bibfnamefont{N.}~\bibnamefont{Gisin}},
  \bibinfo{journal}{Phys. Rev. Lett.} \textbf{\bibinfo{volume}{81}},
  \bibinfo{pages}{3563} (\bibinfo{year}{1998}).

\bibitem[{\citenamefont{Bennett et~al.}(1993)\citenamefont{Bennett, Brassard,
  Cr\'{e}peau, Jozsa, Peres, and Wootters}}]{Bennett:93a}
\bibinfo{author}{\bibfnamefont{C.~H.} \bibnamefont{Bennett}},
  \bibinfo{author}{\bibfnamefont{G.}~\bibnamefont{Brassard}},
  \bibinfo{author}{\bibfnamefont{C.}~\bibnamefont{Cr\'{e}peau}},
  \bibinfo{author}{\bibfnamefont{R.}~\bibnamefont{Jozsa}},
  \bibinfo{author}{\bibfnamefont{A.}~\bibnamefont{Peres}}, \bibnamefont{and}
  \bibinfo{author}{\bibfnamefont{W.~K.} \bibnamefont{Wootters}},
  \bibinfo{journal}{Phys. Rev. Lett.} \textbf{\bibinfo{volume}{70}},
  \bibinfo{pages}{1895} (\bibinfo{year}{1993}).

\bibitem[{\citenamefont{Bennett and Wiesner}(1992)}]{Bennett:92a}
\bibinfo{author}{\bibfnamefont{C.~H.} \bibnamefont{Bennett}} \bibnamefont{and}
  \bibinfo{author}{\bibfnamefont{S.~J.} \bibnamefont{Wiesner}},
  \bibinfo{journal}{Phys. Rev. Lett.} \textbf{\bibinfo{volume}{69}},
  \bibinfo{pages}{2881} (\bibinfo{year}{1992}).

\bibitem[{\citenamefont{Einstein et~al.}(1935)\citenamefont{Einstein, Podolsky,
  and Rosen}}]{Einstein:35a}
\bibinfo{author}{\bibfnamefont{A.}~\bibnamefont{Einstein}},
  \bibinfo{author}{\bibfnamefont{P.}~\bibnamefont{Podolsky}}, \bibnamefont{and}
  \bibinfo{author}{\bibfnamefont{N.}~\bibnamefont{Rosen}},
  \bibinfo{journal}{Phys. Rev.} \textbf{\bibinfo{volume}{47}},
  \bibinfo{pages}{777} (\bibinfo{year}{1935}).

\bibitem[{\citenamefont{Bohm}(1951)}]{Bohm:51a}
\bibinfo{author}{\bibfnamefont{D.}~\bibnamefont{Bohm}},
  \emph{\bibinfo{title}{Quantum Theory}} (\bibinfo{publisher}{Prentice-Hall},
  \bibinfo{address}{New York}, \bibinfo{year}{1951}).

\bibitem[{\citenamefont{Maudlin}(1992)}]{Maudlin:92a}
\bibinfo{author}{\bibfnamefont{T.}~\bibnamefont{Maudlin}}, in
  \emph{\bibinfo{booktitle}{PSA 1992, Volume 1}}, edited by
  \bibinfo{editor}{\bibfnamefont{D.}~\bibnamefont{Hull}},
  \bibinfo{editor}{\bibfnamefont{M.}~\bibnamefont{Forbes}}, \bibnamefont{and}
  \bibinfo{editor}{\bibfnamefont{K.}~\bibnamefont{Okruhlik}}
  (\bibinfo{publisher}{Philosophy of Science Association},
  \bibinfo{address}{East Lansing}, \bibinfo{year}{1992}), pp.
  \bibinfo{pages}{404--417}.

\bibitem[{\citenamefont{Brassard et~al.}(1999)\citenamefont{Brassard, Cleve,
  and Tapp}}]{Brassard:99a}
\bibinfo{author}{\bibfnamefont{G.}~\bibnamefont{Brassard}},
  \bibinfo{author}{\bibfnamefont{R.}~\bibnamefont{Cleve}}, \bibnamefont{and}
  \bibinfo{author}{\bibfnamefont{A.}~\bibnamefont{Tapp}},
  \bibinfo{journal}{Phys. Rev. Lett.} \textbf{\bibinfo{volume}{83}},
  \bibinfo{pages}{1874} (\bibinfo{year}{1999}).

\bibitem[{\citenamefont{Steiner}(2000)}]{Steiner:00a}
\bibinfo{author}{\bibfnamefont{M.}~\bibnamefont{Steiner}},
  \bibinfo{journal}{Phys. Lett. A} \textbf{\bibinfo{volume}{270}},
  \bibinfo{pages}{239} (\bibinfo{year}{2000}).

\bibitem[{\citenamefont{Csirik}(2002)}]{Janos:02b}
\bibinfo{author}{\bibfnamefont{J.~A.} \bibnamefont{Csirik}},
  \bibinfo{journal}{Phys. Rev. A} \textbf{\bibinfo{volume}{66}},
  \bibinfo{pages}{014302} (\bibinfo{year}{2002}).

\bibitem[{\citenamefont{Bacon and Toner}(2002)}]{Bacon:02a}
\bibinfo{author}{\bibfnamefont{D.}~\bibnamefont{Bacon}} \bibnamefont{and}
  \bibinfo{author}{\bibfnamefont{B.~F.} \bibnamefont{Toner}},
\bibinfo{journal}{Phys. Rev. Lett.} \textbf{\bibinfo{volume}{90}},
  \bibinfo{pages}{157904} (\bibinfo{year}{2003}).

\bibitem[{\citenamefont{Cerf et~al.}(2000)\citenamefont{Cerf, Gisin, and
  Massar}}]{Cerf:00a}
\bibinfo{author}{\bibfnamefont{N.~J.}~\bibnamefont{Cerf}},
  \bibinfo{author}{\bibfnamefont{N.}~\bibnamefont{Gisin}}, \bibnamefont{and}
  \bibinfo{author}{\bibfnamefont{S.}~\bibnamefont{Massar}},
  \bibinfo{journal}{Phys. Rev. Lett.} \textbf{\bibinfo{volume}{84}},
  \bibinfo{pages}{2521} (\bibinfo{year}{2000}).

\bibitem[{\citenamefont{Massar et~al.}(2001)\citenamefont{Massar, Bacon, Cerf,
  and Cleve}}]{Masar:01a} 
\bibinfo{author}{\bibfnamefont{S.}~\bibnamefont{Massar}},
  \bibinfo{author}{\bibfnamefont{D.}~\bibnamefont{Bacon}},
  \bibinfo{author}{\bibfnamefont{N.}~\bibnamefont{Cerf}}, \bibnamefont{and}
  \bibinfo{author}{\bibfnamefont{R.}~\bibnamefont{Cleve}},
  \bibinfo{journal}{Phys. Rev. A} \textbf{\bibinfo{volume}{63}},
  \bibinfo{pages}{052305} (\bibinfo{year}{2001}); 
\bibinfo{author}{\bibfnamefont{A.}~\bibnamefont{Coates}}
  (\bibinfo{year}{2002}), \bibinfo{note}{quant-ph/0203112}.

\bibitem[{\citenamefont{Schatten}(1993)}]{Schatten:93a}
\bibinfo{author}{\bibfnamefont{K.~H.} \bibnamefont{Schatten}},
  \bibinfo{journal}{Phys. Rev. A} \textbf{\bibinfo{volume}{48}},
  \bibinfo{pages}{103} (\bibinfo{year}{1993}).

\bibitem[{\citenamefont{Gisin and Gisin}(1999)}]{Gisin:99a}
\bibinfo{author}{\bibfnamefont{N.}~\bibnamefont{Gisin}} \bibnamefont{and}
  \bibinfo{author}{\bibfnamefont{B.}~\bibnamefont{Gisin}},
  \bibinfo{journal}{Phys. Lett. A} \textbf{\bibinfo{volume}{260}},
  \bibinfo{pages}{323} (\bibinfo{year}{1999}).

\bibitem[{\citenamefont{Boschi et~al.}(1998)}]{Boschi:98a}
\bibinfo{author}{\bibfnamefont{D.}~\bibnamefont{Boschi}},
\bibinfo{author}{\bibfnamefont{S.}~\bibnamefont{Branca}},
\bibinfo{author}{\bibfnamefont{F.}~\bibnamefont{De Martini}}, \bibinfo{author}{\bibfnamefont{L.}~\bibnamefont{Hardy}},  \bibnamefont{and}
  \bibinfo{author}{\bibfnamefont{S.}~\bibnamefont{Popescu}},
  \bibinfo{journal}{Phys. Rev. Lett.} \textbf{\bibinfo{volume}{80}},
  \bibinfo{pages}{1121} (\bibinfo{year}{1998}).

\bibitem[{\citenamefont{Braunstein et~al.}(2001)}]{Braunstein:01a}
\bibinfo{author}{\bibfnamefont{S.~L.}~\bibnamefont{Braunstein}},
\bibinfo{author}{\bibfnamefont{C.~A.}~\bibnamefont{Fuchs}},
\bibinfo{author}{\bibfnamefont{H.~J.}~\bibnamefont{Kimble}},  \bibnamefont{and}
  \bibinfo{author}{\bibfnamefont{P.}~\bibnamefont{van Loock}},
  \bibinfo{journal}{Phys. Rev. A} \textbf{\bibinfo{volume}{64}},
  \bibinfo{pages}{022321} (\bibinfo{year}{2001}).

\bibitem[{\citenamefont{Hardy}(1999)}]{Hardy:99a}
\bibinfo{author}{\bibfnamefont{L.}~\bibnamefont{Hardy}},
  \bibinfo{note}{quant-ph/9906123} (\bibinfo{year}{1999}).

\end{thebibliography}

\end{document}